\documentclass[12pt]{iopart}

\usepackage{graphicx}
\usepackage{dcolumn}
\usepackage{bm}
\usepackage{epsfig}
\usepackage{footnote}
\usepackage{times}

\begin{document}
\title{Practical long-distance quantum key distribution system using decoy levels}

\author{D Rosenberg$^1$, C G Peterson$^1$, J W Harrington$^1$, P R Rice$^1$, N Dallmann$^1$, K T Tyagi$^1$, K P McCabe$^1$, S Nam$^2$, B Baek$^2$, R H Hadfield$^3$, R J Hughes$^1$ and J E Nordholt$^1$}
\address{$^1$Los Alamos National Laboratory, Los Alamos, NM}
\address{$^2$National Institute of Standards and Technology, Boulder, CO}
\address{$^3$Heriot-Watt University, Edinburgh, United Kingdom}

\begin{abstract}
Quantum key distribution (QKD) has the potential for widespread
real-world applications.  To date no secure long-distance experiment
has demonstrated the truly practical operation needed to move QKD
from the laboratory to the real world due largely to limitations in
synchronization and poor detector performance. Here we report
results obtained using a fully automated, robust QKD system based on
the Bennett Brassard $1984$ protocol (BB$84$)\cite{BB84} with
low-noise superconducting nanowire single-photon detectors (SNSPDs)
and decoy levels.  Secret key is produced with unconditional
security over a record ${144.3}$~km of optical fibre, an increase of
more than a factor of five compared to the previous record for
unconditionally secure key generation in a practical QKD system.
\end{abstract}



\section{Introduction}
Quantum key distribution (QKD) holds the promise of communication
with security resting firmly on the foundation of quantum mechanics
rather than unproven assumptions regarding current and future
computational resources.  In the BB$84$ protocol \cite{BB84}, the
most common prepare and measure protocol, the sender (Alice) ideally
encodes a single photon with a bit value of either $0$ or $1$ in one
of two conjugate bases and sends it to the receiver (Bob). When Bob
receives a photon, he measures it in one of the two bases. Alice and
Bob then publicly share their basis choices and only keep the bits
where the bases match.  Because the information is encoded in a
single photon, any tampering by an eavesdropper (Eve) results in an
increased error rate detectable by the legitimate users of the
system.  After performing error correction \cite{Brassard94} to
remove any errors that may have arisen from the operations of an
eavesdropper or from imperfections in the experimental apparatus,
Alice and Bob perform a privacy amplification step \cite{Bennett95}
to erase any partial information Eve might have obtained about the
transmission.  The final result is a string of $0$s and $1$s that
Alice and Bob share but about which Eve has negligibly small
information.

Although it is straightforward to see how QKD with ideal sources and
detectors is secure, real-world QKD systems must contend with
imperfect experimental components. In $2000$, it was pointed out
that the use of attenuated lasers (which emit photons in a
Poissonian distribution) drastically limits the secure range of
laser-based QKD systems \cite{Brassard00}. Eve could perform a
photon number splitting (PNS) attack in which she selectively
changes the channel transmittance based on photon number but still
maintains the expected rate of detections at Bob. If the rate of
multi-photon pulses at Alice is larger than the rate of photon
detections at Bob, then Eve could block all the laser pulses
containing single photons, and only allow transmission of the
multi-photon pulses, which are inherently insecure. In this case,
the entire string of ``secure'' bits could, in fact, be known to
Eve.

Several methods have been proposed to mitigate the effects of PNS
attacks. The most straightforward is to simply use a very low mean
photon number $\mu$, on order of the channel transmittance, to
guarantee that at least some of the detections at Bob originated
from single photons at Alice \cite{Brassard00}. Unfortunately, this
results in very low secret bit rates as well as greatly limiting the
possible transmission length due to dark counts in the detectors.
The SARG protocol \cite{SARG}, which allows for secure bits to be
formed from both one- and two-photon signals, outperforms standard
BB$84$ , but it offers no performance advantage compared to BB$84$
with ``decoy states'' as described below \cite{SARGvBB84}.
Differential phase shift QKD (DPS-QKD) is another method of
protecting against PNS attacks \cite{Takesue}, but there does not
currently exist a security proof against all attacks for the DPS-QKD
protocol, so it does not provide the unconditional security needed
for QKD.

The use of decoy states \cite{Hwang_decoy,Lo05,Wang05,Harrington05}
with the BB$84$ protocol has provided a method for achieving high
bit rates while protecting against PNS attacks and maintaining the
underlying security of BB$84$. In a decoy state protocol, Alice
transmits signals randomly picked from several different mean photon
levels $\mu_j$ rather than just one. If Eve, who is ignorant of the
$\mu_j$ value for each specific signal, were to attempt to perform a
PNS attack, she would not be able to simultaneously modify the
channel transmission for all the $\mu_j$ values to reproduce the
expected statistics at Bob. By comparing the number of photon
detections at each mean photon level, Alice and Bob can determine a
rigorous bound on the number of single photons that have been
received by Bob and incorporate this bound into the privacy
amplification step.

Decoy state protocols have been demonstrated over a free-space link
\cite{freespacedecoy_07} and in several different fibre systems
\cite{Lo_15km,Lo_60km,PRL07,chinadecoy_07,Shields_APL07,Shields_optexp07,China_decoy_123km,Hasegawa07}.
However, many of the demonstrations, although they are important
proof-of-principle experiments, do not perform the full QKD
protocol, including error correction and privacy amplification, but
instead only estimate the secret bit rate and range of the system.
Furthermore, most systems have employed either local synchronization
(transmitter and receiver share the same clock) or synchronization
over a separate fibre, methods that pose a significant barrier for
deployment. Although a practical uni-directional system with remote
synchronization has been demonstrated over short distances of
$25$~km with high bit rates \cite{Shields_APL07,Shields_optexp07},
the detectors used in these experiments were not sufficiently quiet
to enable long-distance fibre QKD. Here we describe the first
practical QKD system capable of secure operation over distances
greater than $100$~km. The system uses a three-level decoy-state
protocol in a fibre phase-encoding quantum key distribution system
with independent clocks at the transmitter and receiver and
low-noise superconducting nanowire single photon detectors (SNSPDs).
In addition to being secure against PNS attacks, our data is secure
against a number of other attacks, including Trojan Horse attacks
that affect bi-directional systems, timing attacks based on
exploiting detector efficiency mismatch \cite{Lo_timing07}, and
attacks based on differing timing responses between detectors
\cite{Kurtseifer_timing07}.

\section{Finite statistics decoy state protocol}

The security of our protocol rests on knowing the number of sifted
bits arising from single photons produced by Alice, and the
disturbance, or bit error rate, on those bits.  Our security
statement incorporates finite statistics into each step of the
protocol. The number of secret bits distilled from each basis is
calculated as a modification from \cite{GLLP}
\begin{eqnarray}\label{nsec_eq}
\!\!\!\!\!\!\!\!\!\!\!\!\!\!\!\!\!  
N_{\rm secret} & = & N_{\rm sifted} \left[ y_1^- \mu e^{-\mu} \left(
1 - f_{\rm PA} H_2(b_1^+) \right)
- f_{\rm EC} H_2(B) - \left(1 - \frac{1}{f_{\rm DS}} H_2(z) \right)
\right]
\end{eqnarray}
where $N_{\rm sifted}$ is the number of sifted bits in the selected basis,
$B$ is the observed bit error rate in this basis, $y_1^-$ is a lower bound
on the transmittance of single photons, $b_1^+$ is an upper bound on
the single-photon bit error rate in the conjugate basis, $z$ is the fraction
of zeroes among the sifted bits in this basis, and $H_2(\cdot)$ is the
Shannon binary entropy function, and $f_{\rm PA}$, $f_{\rm EC}$, and
$f_{\rm DS}$ are privacy amplification, error correction, and deskewing
efficiency factors needed to accommodate the finite statistics of the data.

The decoy state protocol allows us to determine a lower bound on the
single photon transmittance $y_1^-$ and an upper bound on the single
photon error rate $b_1^+$. The value of $y_1^-$ is the minimal value
of $y_1$ satisfying the following inequalities:
\begin{eqnarray*}
Y_j^- ~\leq~ e^{-\mu_j} \sum_{n=0}^{\infty}
\frac{\left(\mu_j\right)^n}{n!}y_n ~\leq~ Y_j^+
\end{eqnarray*}
where $y_n$ is the transmittance of an $n$-photon signal and $Y_j^+$
($Y_j^-$) is the upper (lower) bound on the yield of detection
events when mean photon number $\mu_j$ is used for transmission.  In
contrast to other work where $Y_j^\pm$ are calculated assuming that
the underlying detection statistics are Gaussian, we make no such
assumption and use the full binomial distribution to calculate the
bounds within some user-defined confidence level, chosen to be
$1\times 10^{-7}$ for this experiment.

We use two methods to determine the single photon error rate $b_1$.
The first, referred to as the ``worst-case'' scenario, makes the
conservative assumption that all observed errors occur on single
photons. In this case, $b_1^+$ is simply equal to the number of
observed errors divided by the number of sifted bits that arose from
single photons prepared by Alice, which can be computed from the
lower bound $y_1^-$.  However, we can obtain a tighter bound on
$b_1$ by utilizing the information contained in the differing error
rates at each $\mu_j$ and constructing a set of inequalities
involving the bounds on the observed bit error rates $B_j$ and the
$n$-photon bit error rates $b_n$:
\begin{eqnarray*}
B_j^- ~\leq~ e^{-\mu_j} \sum_{n=0}^{\infty}
\frac{\left(\mu_j\right)^n}{n!}y_n b_n ~\leq~ B_j^+
\end{eqnarray*}
Analogous to the method for finding $y_1^-$, $b_1^+$ is found by
determining the maximal value of $b_1$ that satisfies these
inequalities.  Computing this tighter bound on $b_1$ can be carried
out by linear programming, as for $y_1^-$, but here it is subject to
quadratic constraints, so the analysis is computationally more
intensive, although still tractable.  In either case, the bounds on
$y_1$ and $b_1$ are valid even if the quantum channel is
time-varying.

Equation \ref{nsec_eq} involves three efficiency factors ($f_{EC}$,
$f_{PA}$ and $f_{DS}$) which quantify finite statistics effects
during the stages of QKD. Asymptotically, all three factors approach
unity, but for a finite session length they are strictly greater
than one.  Reconciliation via practical error-correcting algorithms
does not achieve the Shannon capacity of the binary symmetric
channel, and the extra overhead in parity checks that needs to be
communicated between Alice and Bob is expressed by the factor
$f_{EC}$. Privacy amplification involves removing any partial
information Eve may have gained by disturbing those single-photon
signals present among the sifted bits.  Following Koashi
\cite{Koashi05}, we numerically compute the logarithm (base two) of
the number of typical strings that are needed to describe with high
confidence the output of a binary symmetric channel with a given bit
flip probability, using the bit error rate in one basis to determine
the amount of privacy amplification required in the other basis. The
factor $f_{PA}$ denotes how much the size of this output differs
from the Shannon entropy of the single-photon signals.
The last finite statistics effect we consider is due to the
imbalance between the two detector efficiencies, which leads to a
bias between the zeroes and ones for the sifted bits in each basis.
This bias can reduce Eve's search space of guessing over likely
reconciled keys prior to privacy amplification. Asymptotically,
Shannon entropy again gives the required reduction in secret
information, but any practical algorithm for removing the bias, or
{\it deskewing}, is likely to have inefficiencies, which is
encompassed by the factor $f_{DS}$. We followed Peres \cite{Peres}
in iterating von Neumann's algorithm for generating unbiased bits
out of the reconciled keys to determine the value of $f_{DS}$.

\section{Automated QKD system}

\begin{figure}
  \begin{center}
  \includegraphics[width=5 in]{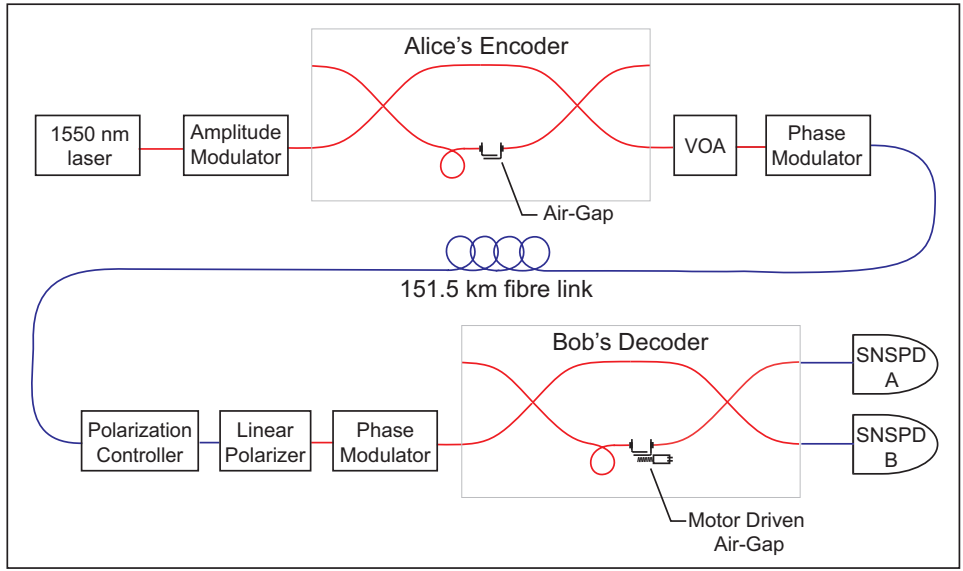}
  \caption{\label{system_diagram_fig} Optical train of the
phase-encoding quantum key distribution system used in this work.
The variable optical attenuator (VOA) provides attenuation to the
single photon level, from which the amplitude modulator provides the
different $\mu_j$ levels needed for the decoy level protocol. Red
lines indicate polarization maintaining fibre, and blue lines
represent single mode fibre. Photons are detected by two
superconducting nanowire single photon detectors (SNSPDs) biased to
achieve matching detection efficiencies.}
  \end{center}
\end{figure}

A diagram of the automated, reconfigurable QKD system used
\cite{Hughes05} is shown in Figure~\ref{system_diagram_fig}. A
$1550$~nm distributed feedback laser is pulsed at a clock rate of
$10$~MHz, and the resulting photons are sent to Alice's phase
encoder. The photon wavepacket is split into a ``short'' and a
``long'' path, and the portion of the wavepacket that traversed the
long path is modulated by the electro-optic phase modulator, located
outside the interferometer for stability. Likewise, the wavepacket
is again split in Bob's decoder, and only the part of the wavepacket
that traveled through Alice's long path is modulated. Interference
ultimately occurs between the Alice-long-Bob-short and
Alice-short-Bob-long amplitudes. Polarization maintaining fibre is
used within the interferometers to ensure that these two paths are
indistinguishable.

The quantum channel consists of $151.5$~km of optical fibre with an
attenuation of $.206$~dB per km, and shorter distances are obtained
by redefining Alice's enclave to include a portion of the fibre. The
bit and basis selections at Alice and Bob are all determined by the
output of physical random number generators, and a high-speed
optical switch, driven by a pseudo-random pattern, provides the
three intensity levels needed for our decoy state protocol. (In a
deployed system, the pseudo-random pattern generator would simply be
replaced by a physical random number generator.) The superconducting
nanowire single photon detectors (SNSPDs) \cite{Hadfield05} used for
these measurements are cooled to $3$~Kelvin in a closed-cycle
refrigeration system and coupled to single-mode telecom fibre. They
operate in ungated mode with a measured full width at half maximum
timing jitter of $69$ ps and a recovery time of $<10$ ns. The
detectors were individually current biased to a matching detection
efficiency of $0.5$\%, resulting in a summed average dark count rate
over the entire acquisition period of $78.1$ counts/second.

The system is fully automated and able to run for many hours without
user intervention. Independent Rubidium oscillators are employed as
frequency references at Alice and Bob and are synchronized using
only the quantum signals.  For each acquisition period (typically
$1$--$10$ seconds), a photon arrival time histogram is created that
is used to determine the current average frequency offset between
the two oscillators, which is corrected by adjusting one of the
clocks' frequencies. Long-term system stability is achieved through
a combination of tuning and QKD runs. Tuning runs are insecure
sessions executed at a high photon number to quickly obtain the
statistics needed for interferometer and timing adjustments, while
QKD runs adhere to the requirements for secure key generation. To
minimize the effects of system de-tuning, tuning runs are promptly
followed by QKD runs. The use of ungated detectors and
post-selection of detection timestamps in software relaxes the
system timing requirements to the phase modulator voltage pulse
width, $2$--$3$~ns, rather than being constrained to the
sub-nanosecond electrical detector gate width required for avalanche
photodiodes. Periodically, the polarization controller is
automatically adjusted to compensate for any polarization drifts
within the fibre by maximizing transmission through the linear
polarizer.

\section{Results and conclusions}
We chose $\mu_j$ values with associated sending probabilities
that were near-optimal for a fibre length of $135$~km.
The selected mean photon numbers were
$\left\lbrack\mu_0 = 0.0025, \mu_1 = 0.13, \mu_2 = 0.57\right\rbrack$
($\mu_0$, ideally the vacuum state, was constrained to be $23.5$~dB
below the high $\mu_2$ level by the extinction ratio of the switch) with
associated sending probabilities $\left\lbrack 0.1, 0.2, 0.7 \right\rbrack$.
In $5.6$ hours of acquisition time and using a timing window of $184$~ps,
the number of detection events recorded at $\mu_2$, $\mu_1$, and $\mu_0$
was $80776$, $5729$, and $341$, respectively, and a total of $40538$
sifted bits with a bias (fraction of zeros in one basis) of $0.494$
were created. The sifted bits and the basis choices both passed the
FIPS $140$-$2$ cryptographic randomness tests \cite{FIPS}. After data
were collected, the bits were sifted, error corrected using the
modified CASCADE algorithm \cite{modCASCADE}, and privacy amplified
as described in equation (\ref{nsec_eq}) after determining the
bounds on the single photon transmittance and error rate. The
efficiency factors due to the finite data size were determined to be
$f_{EC} \approx 1.07$, $f_{PA} \approx 1.09$ and $f_{DS} \approx
1.05$.

\begin{figure}
  \begin{center}
  \includegraphics[width=5in]{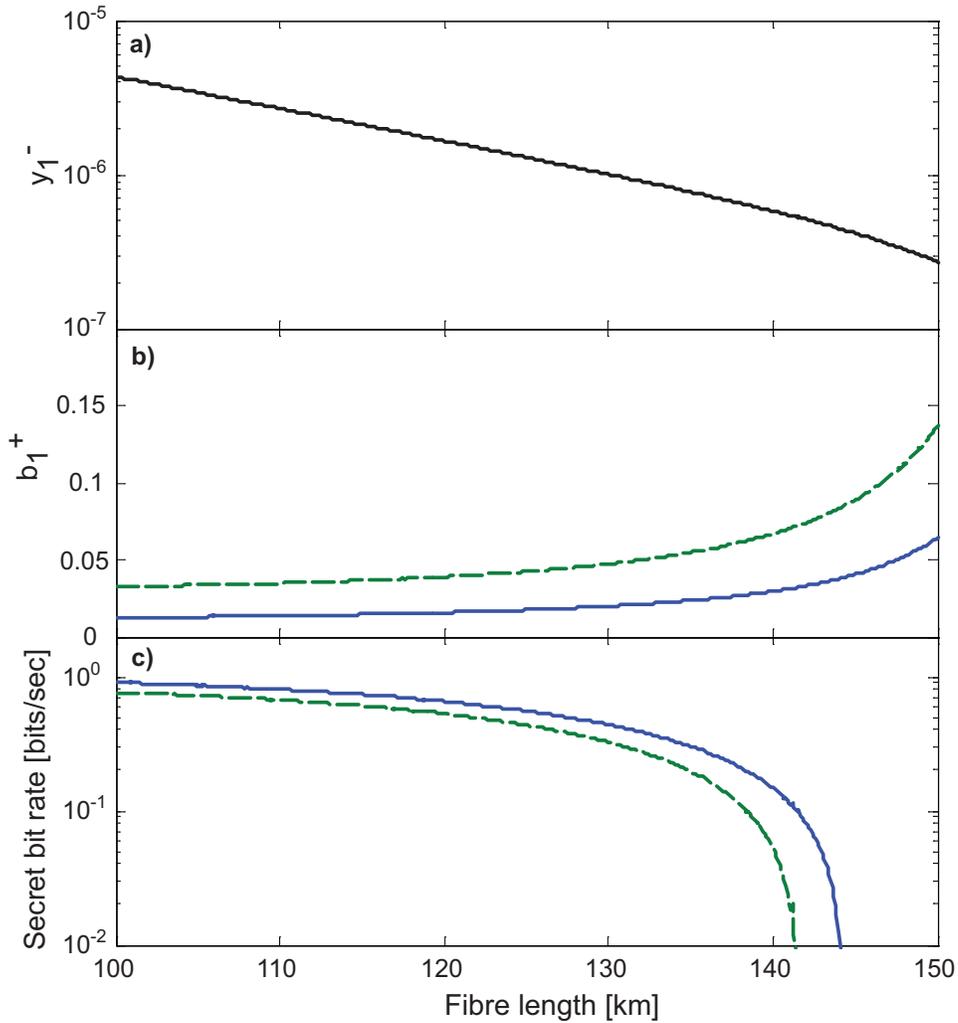}
  \caption{\label{longdist_fig} (a) Bound on single photon transmittance
  $y_1^-$; (b) bounds on the single photon error rate $b_1^+$ assuming either
  the ``worst-case'' (dashed green line) or by finding a tighter bound
  (solid blue line); and (c) secret bit rate as a function of distance.
  Data was acquired over a $151.5$~km link, and the decoy levels and
  sending probabilities were chosen to maximize the secret bit rate
  at $135$~km.  }
  \end{center}
\end{figure}

As shown in Figure \ref{longdist_fig}, $b_1^+$, the bound on the
single photon error rate, is significantly higher when the
worst-case assumption that all the errors occur on single photons is
made. The tighter $b_1$ bound not only yields more secret bits at
any given distance, but it also extends the distance over which
secret bits can be exchanged.  Use of the ``worst-case'' $b_1$
results in $3990$ secret bits at $135$~km and a maximum range of
$141.6$~km.  When the tighter bound on $b_1$ is used, $6127$ secret
bits are produced at $135$~km and the range of the system is
extended to $144.3$~km, a new record for key distribution secure
against general attacks \cite{GLLP}, including
photon-number-splitting attacks.

\begin{figure}
  \begin{center}
  \includegraphics[width=5in]{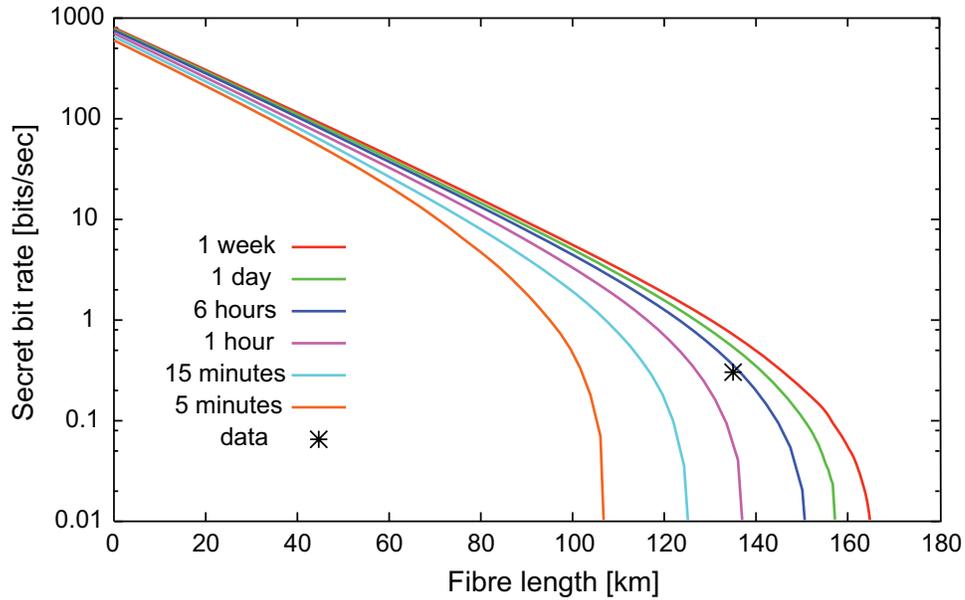}
  \caption{\label{simf3a_fig} Simulated performance of system used
in this work for different acquisition times and distances. The
optimal mean photon number and sending probabilities are found for
each distance and acquisition time.  The data point at $135$~km
matches the prediction of the simulation.  Even with an efficiency
of $0.5\%$, the SNSPDs enable the creation of secret bits out to
$166.1$~km.}
  \end{center}
\end{figure}

By choosing different mean photon numbers and sending probabilities
and acquiring data for longer times, it would be possible to extend
the range even further in this system.  Figure \ref{simf3a_fig}
shows the results of simulations based on the system properties to
determine the maximum range of the system.  With the same detectors
used in this work, this system could achieve a range of $166.1$~km.
Although the detection efficiency of these detectors is currently
quite low, by embedding the detectors in a stack of optical elements
designed to increase absorption at the target wavelength
\cite{SSPD_higheff} and improving the optical coupling to the
detector, it should be possible to increase the system detection
efficiency considerably. An increase to $50\%$ (assuming the same
dark count and background photon rates) would result in bit rates
higher by approximately two orders of magnitude at any given
distance in the asymptotic limit, and an increase in the tolerable
link loss to over $50$~dB. From our results, which are the first
demonstration of assured security over long distances in a
deployable system, we can infer that reasonable improvements in
component technology will result in several hundred bits per second
over distances of $100$~km, more than an order of magnitude higher
than what is presently available.


\ack
 The authors thank Joe Dempsey and Corning, Inc. for the optical
 fibre, Thomas Chapuran and Nicholas Peters for helpful discussions,
 and G. Gol'tsman for providing the original detectors used in this
 work.  The authors acknowledge support from IARPA, the Department
 of Commerce, the NIST quantum information science initiative, and
 the Royal Society of London.

\end{document}